\documentstyle[german,psfrag,titlepage,epsf,amsmath,a4paper,15pt]{afi}
\makeindex
\begin{document}

\newcommand{\bfm}{\boldsymbol}
\def\partialt{\partial_t}
\def\bnab{\bnabla \! \!}
\def\bb{{\bf B}}
\def\simgeq{\mathrel{\smash{\mathop{\raise2pt\hbox{$>$}}\limits_%
   {\smash{\raise4pt\hbox{$\sim$}}}}\vphantom\geq}}

\def\@abstract[#1]{%
  \global\@hasabstracttrue
  \hyphenpenalty\sv@hyphenpenalty     % restore \hyphenpenalty
  \global\setbox\t@abstract=\vbox\bgroup
  \leftskip\z@
  \@rightskip\z@ \rightskip\@rightskip \parfillskip\@flushglue
   \@abstractsize                      % Text in 9/11
  \parindent 1em                      % \parindent in abstract
  \noindent {\bfseries\abstractname}  % caption `Abstract' (bold)
  \vskip 0.5\@bls    % half a line of space below
\noindent\ignorespaces
}

\pagenumbering{roman}
%\include{Titel}
%\newpage

%{\iupsize \baselineskip 17pt \tableofcontents}

%\skippage
\pagenumbering{arabic}
\setcounter{page}{1}
%\setcounter{page}{5}

%\skippage

\iupsize
\baselineskip 18pt

\chapter{\noindent
The evidence for unusual gravity 
from the large-scale structure of the Universe
}

%\vspace{0.5cm}

A. Diaferio 

\vspace{0.5cm}

\noindent{\normalsize \it
Dipartimento di Fisica Generale ``Amedeo Avogadro'', Universit\`a degli Studi 
di Torino, and 
Istituto Nazionale di Fisica Nucleare, Sezione di Torino, Torino, Italia}

\vspace{0.5cm}

\noindent{
\normalsize
Under the assumption that General Relativity (GR) correctly describes the phenomenology
of our Universe, astronomical observations provide compelling evidence that
(1) the dynamics of cosmic structure is dominated by dark matter (DM), an exotic
matter mostly made of hypothetical elementary particles,
and (2) the expansion of the Universe is currently accelerating
because of the presence of a positive cosmological constant $\Lambda$.
The DM particles have not yet been detected and there is no theoretical justification
for the tiny positive $\Lambda$ implied by observations. Therefore,
over the last decade, the search for extended or alternative theories of gravity
has flourished.  
}

\vspace{0.5cm}

\section{The evidence for dark matter and $\Lambda$}

\subsection{Systems of galaxies}

The first evidence of the existence of DM was found by Zwicky \cite{zwicky33,zwicky37}.
He measured the redshift of the eight brightest galaxies in the
Coma cluster, attributed their redshifts to the Doppler effect, and assumed 
the cluster to be in virial equilibrium. 
The galaxy velocity dispersion thus promptly gives the total
mass of the cluster, that turns out to be
$\sim 100$ times larger than the sum of the masses
of the individual galaxies. The obvious way out, that the cluster
has a positive total energy, namely that it is expanding, is not convincing, because 
the cluster should have just formed and if this were the 
case for the many clusters we see on the sky, it would be unlikely to observe
them.
The cluster could also be a chance fluctuation in the galaxy distribution,
but this hypothesis is called into question by the morphological
and photometric homogeneity of the galaxy population of the cluster.
In the following decades, this very same problem also appeared in 
nearby groups of galaxies \cite{amba61}.

In the early 1970s, the measure of the 21-cm line emission of neutral hydrogen 
showed that the rotation curves of spirals do not fall off
at large radii, as expected if most of the galaxy mass were
concentrated in the optically luminous component \cite{roberts73}
(see \cite{salucci07} for a recent review). 
These findings appeared to be essential to explain the 
stability of the disk of spiral galaxies: in fact, self-gravitating
disks, which are prone to bar instabilities, become dynamically stable when they
are embedded within massive halos \cite{ostriker73, einasto74}.

At the same epoch, clusters of galaxies were discovered to be
strong X-ray emitters \cite{gursky72}. If the emission mechanism is thermal, due to
line emission or bremsstrahlung which originate in the ion-ion and ion-electron collisions
in the intracluster plasma, the plasma temperature is an indicator
of the depth of the gravitational potential well. The derived mass of the
cluster agrees, within a factor of two in most cases, with the traditional estimate based 
on the virial theorem or the Jeans equations which only use kinematic data 
\cite{biviano06}. Moreover, the very existence of the
intracluster  plasma is a strong indication that there must be a massive
halo that keeps the plasma confined. In 1990s, hot plasma
was also observed in many groups (Figure \ref{fig:ngc2300}), 
showing that they roughly are a rescaled version
of clusters \cite{mulchaey93}.  

\begin{figure}[t]
        \centerline{\epsfxsize=0.5\hsize \epsffile{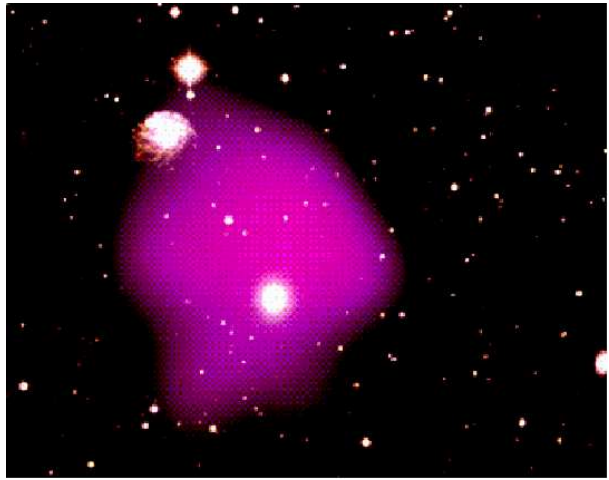}}
        \caption[*]{Diffuse X-ray emission in the NGC 2300 group of galaxies, 
also known as Hickson compact group HG92. From \cite{mulchaey93}.
The presence of a confined X-ray emitting intergalactic plasma indicates
that the compact group is gravitationally bound \cite{diaf95}.
\label{fig:ngc2300}}
\end{figure}

Both the galaxy kinematics and the X-ray analysis assume that the
cluster is in virial equilibrium. In 1990s, photometry
reached enough accuracy that one could start using clusters as gravitational
lenses: the images of background galaxies are distorted
by the gravitational potential well of the cluster and the cluster mass
can be inferred from the amount of this distortion \cite{kaiser93}. 
This technique does not need any assumption about 
the dynamical state of the cluster. Another mass-estimation technique 
which does not assume virial equilibrium, known as the caustic method,
was developed by Diaferio and Geller \cite{diaferio97}; this technique is
only based on the distribution of galaxies in redshift space.
These caustic and gravitational lensing methods,
which are based on completely independent physical principles and data, provide 
cluster masses in surprisingly good agreement. They also agree
with the traditional kinematic and X-ray methods in many cases. When they disagree
with these two latter traditional methods, the caustic and lensing methods still agree 
with each other: 
this result is a clear indication that in these cases the cluster is out of
equilibrium \cite{diaferio01}.

In 1965, Abell made the argument that the mean mass density of the Universe must
be smaller than the mean density within clusters and larger than the mean density
computed including only the richest clusters \cite{abell65}. He thus estimated 
a density of the Universe $\Omega_m\approx 0.2$,
which is remarkably close to the currently accepted value $\Omega_m=0.26$ \cite{spergel07}.
This value implies a cosmic mass-to-light ratio $\langle M/L_B\rangle = \Omega_m 
\rho_c/\rho_L \simeq 400 h $ M$_\odot $~L$_\odot^{-1}$, 
where $\rho_c=3H_0^2/8\pi G$ is the critical density
of the Universe and $\rho_L\sim 1.7\times 10^8 h$ L$_\odot$~Mpc$^{-3}$ 
is the luminosity density of the Universe derived from the galaxy luminosity
function. Including the contribution of their DM halo, 
galaxies have a typical mass-to-light ratio $\langle M/L_B\rangle
\sim 10 h $ M$_\odot$~L$_\odot^{-1}$. Therefore, galaxies globally contribute 
less than 3\% to the mass in the 
Universe and we must conclude that DM is mostly distributed in
structures of size much larger than the size of galaxies.

\subsection{The large-scale structure}   

\begin{figure}[t]
        \centerline{\epsfxsize=0.8\hsize \epsffile{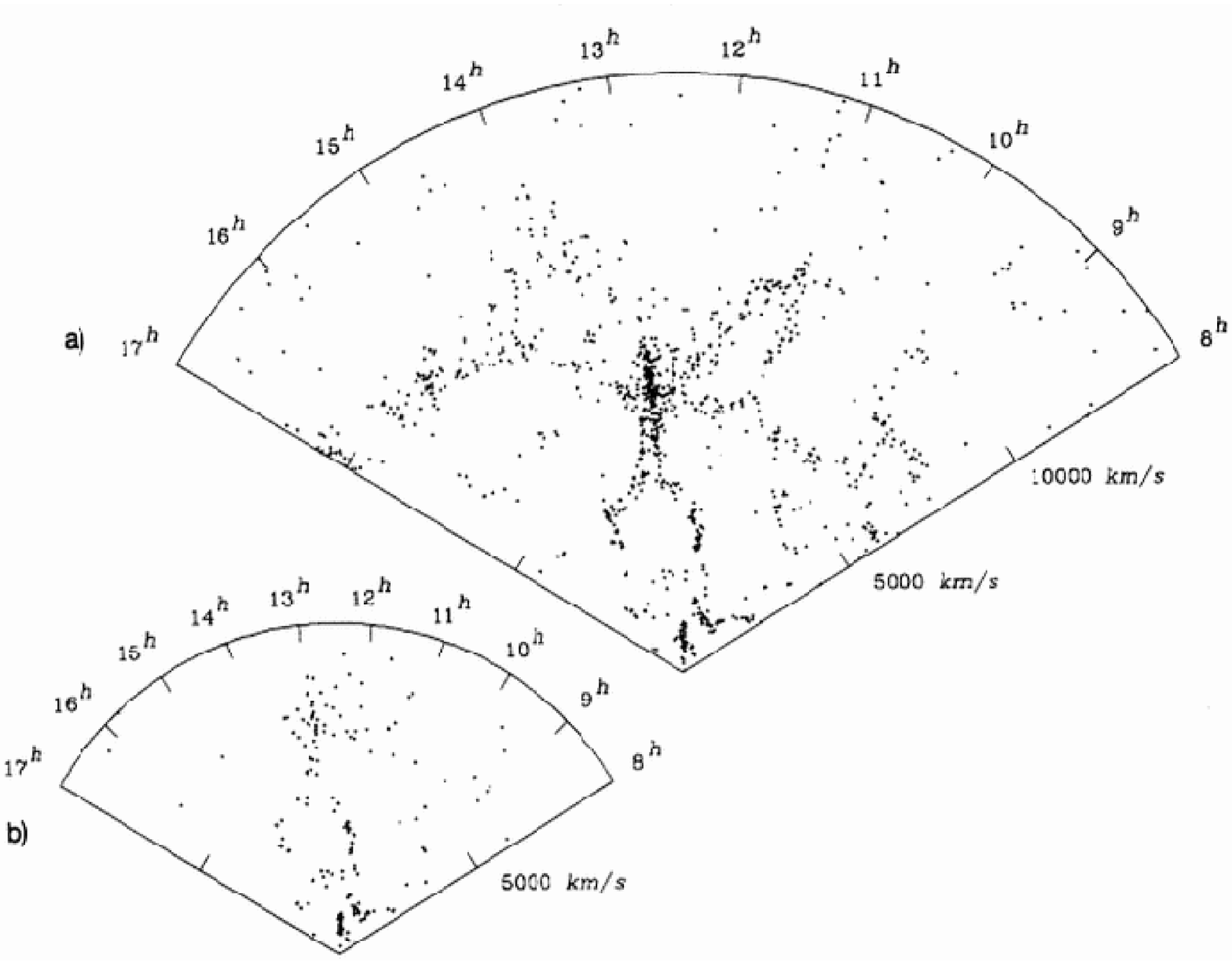}}
        \caption[*]{Each dot is a galaxy in the wedge with 
declination in the range $26^\circ.5-32^\circ.5$ and right ascension and redshift
$cz$ as shown. The upper panel shows the 1061 galaxies with $m_B<15.5$ and $cz<15,000$ km s$^{-1}$ 
and the lower panel the 182 galaxies with $m_B<14.5$ and $cz<10,000$ km s$^{-1}$. 
From \cite{delapp86}.
\label{fig:cfa}}
\end{figure}

In 1980s, advances in the optical detector technology made possible the realization
of large redshift surveys. The most striking result
appeared in 1986 when de Lapparent, Geller and  Huchra \cite{delapp86}
published the first slice of the extension of the Center for Astrophysics redshift
survey \cite{davis82}: 
with $\sim 1100$ galaxies in a $6^\circ\times 117^\circ$ strip on the sky and a depth 
of 15,000 km s$^{-1}$ in redshift, this survey imposed the view, contrary to
what was commonly believed at that time, that the distribution
of galaxies is largely dishomogeneous on scales of tens of megaparsecs (Figure \ref{fig:cfa}).
More recent surveys \cite{colless01, york00}  confirmed this result.
Currently, the largest survey is the SDSS with $\sim 700,000$ galaxy redshifts.

How does this large cosmic structure form? Gravitational instability
is the simplest driving process we can think of. In 1981, the inflationary
scenario was proposed to solve the three classical problems
of the standard hot Big Bang cosmology: the horizon, the flatness and
the magnetic monopole density \cite{guth81}. The inflationary scenario
also gives, as a bonus, the initial conditions of the matter
density field: small perturbations due to quantum fluctuations
are inflated to cosmic scales by the tremendous expansion 
of the Universe \cite{mukh81}.

Superclusters of galaxies typically have matter overdensities
$\delta$ of the order of $\sim 1-10$. Gravitational instability
yields a growth rate $\propto (1+z)^{-1}$, with redshift $z$, or slower. Thus, 
the temperature anisotropies $\delta T/T= \delta/3$ in the Cosmic Microwave
Background (CMB), which formed at $z\sim 10^3$, should be larger than one to
yield the cosmic structure we see today on large scales.
On the contrary, observations yield $\delta T/T \sim 10^{-5}$ on $\theta \sim 7^\circ$
angular scales, corresponding to superclusters and larger structures \cite{smoot92}. 
To reconcile these tiny CMB anisotropies with
the gravitational instability paradigm, we need to assume
that the DM density perturbations start evolving at the
time of equivalence, when the density perturbations 
in the baryonic matter\footnote{In the astronomical jargon, 
baryonic or ordinary matter is all the matter made of quarks and electrons,
although, rigorously, only strongly interacting fermions are baryons. 
All the other matter is generically called non-baryonic matter, including
neutrinos and the hypothetical Weakly Interacting Massive Particles (WIMPs).}
are still coupled to the
radiation field. The baryonic perturbations 
fall later into the DM gravitational potential wells when,
at recombination time, they find a dishomogeneous distribution of the
dynamically dominant DM. Therefore,
we are forced to assume the existence of a non-baryonic DM, 
that can decouple much earlier than the baryons 
responsible for the CMB anisotropies. 

Further evidence of the non-baryonic nature of DM is
the abundance of light elements which are synthetized
in the early Universe. Measures of the primordial abundance of deuterium,
for example, which is particularly sensitive to the photon-to-baryon ratio,
implies a baryon density $\Omega_b h^2=0.0214\pm 0.0020$ \cite{kirkman03}, which
is sensibly smaller than the total matter density $\Omega_m= 0.26$, 
and agrees with the baryon density  $\Omega_b h^2=0.02229\pm 0.00073$ 
implied by the CMB anisotropies \cite{spergel07}.

It was soon clear that DM cannot be mostly
made of neutrinos, because the large-scale distribution would
be much fuzzier than observed \cite{white83},
and cosmic structure would have
formed in a top-down rather than in a bottom-up fashion, as indicated, for example,
by the existence of quasars and galaxies at high redshift \cite{fan04}.
Therefore, DM must be mostly made of
cold collisionless particles (CDM), namely 
particles that were not relativistic at the time of
decoupling \cite{davis85}.

These conclusions can be reached by exploiting the fact that DM dominates the dynamics 
of the large-scale structure, and we can, to first approximation, 
neglect the complications due to the dissipative
nature of baryons on scales larger than galaxy clusters.
However, a more detailed comparison with the real Universe requires the
modeling of galaxy formation. Various methods at
different level of sophistication show that a CDM scenario reproduces
the evolution of kinematic, photometric and clustering properties of galaxies 
\cite{kauff99, springel05}. 
Discrepancies remain \cite{schmalz00, casa06}, but they are likely to be due to the
approximated galaxy formation recipes rather than to an incorrect DM modelling.

To partially avoid the difficulty of baryon physics, 
we can resort to weak gravitational lensing: the light emitted by galaxies at
high redshift reaches our telescopes after being 
deflected by the dishomogeneities of the DM distribution which 
increase with time \cite{kaiser92}.
Thus, the weak lensing analysis probes both the amount of DM and the history
of structure formation.

The validity of the CDM scenario obtained a relevant success with 
quasar spectra: the short-wavelength side of these spectra contains the 
Lyman-$\alpha$ forest, a large number of absorption
lines. These lines were commonly attributed to 
clouds of neutral hydrogen between the quasar and the observer. N-body/hydrodynamical
simulations of a CDM universe, initially not conceived to reproduce 
the statistical properties of these absorption lines,
indeed produced synthetic spectra of quasars in amazingly
good agreement with observations, indicating
that the Lyman-$\alpha$ ``cloud'' interpretation was unnecessarily elaborate 
\cite{hernquist96}.

\begin{figure}[t]
        \centerline{ \epsfxsize=0.6\hsize
\epsffile{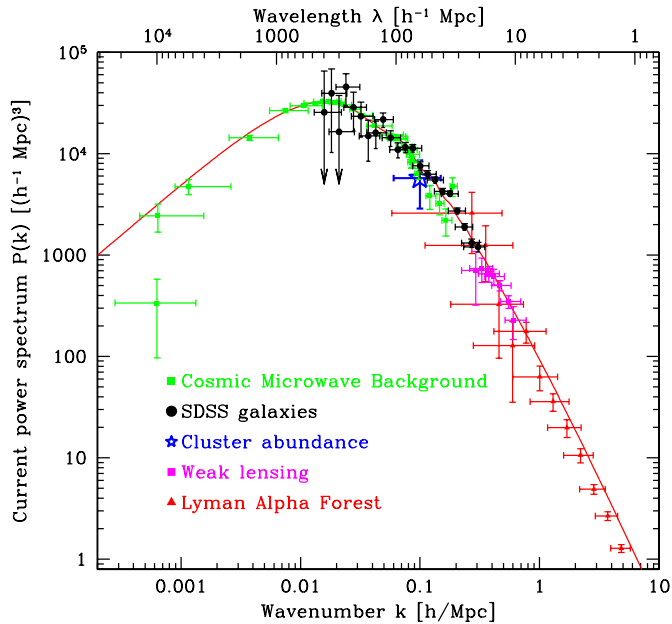}}
        \caption[*]{
Large-scale structure power spectrum for a 
flat scalar scale-invariant CDM model with $
\Omega_m=0.28$, Hubble constant $h=0.72$, and $\Omega_b=0.0448$,
assuming a bias parameter $b=0.92$ for the SDSS galaxies. From \cite{tegmark04}.
\label{fig:Pk}}
\end{figure}

When all this information on the matter distribution on large scales,
CMB, galaxy distribution,
cluster abundance and dynamics, weak lensing and Lyman-$\alpha$ forest measures,
is combined into a power spectrum of the large-scale structure
density fluctuations, we find that the best fit is obtained with a non-null
cosmological constant $\Lambda$ (Figure \ref{fig:Pk}) \cite{tegmark04}.  
Therefore, the cosmological community was not too surprised when, at the end of last century, 
some high-$z$ supernovae were observed to be fainter than expected in a decelerating universe,
namely a universe with a null cosmological constant \cite{riess98, perlmutter99}. 
More recent supernova samples confirm that $\Lambda$ must be positive,
but both the statistical and systematic uncertainties remain large \cite{cloc06}.

In conclusion, cosmological observations 
lead us to a suspicious cosmic energy budget: baryon mass density $\Omega_b\sim 0.04$, 
non-baryonic dark matter density $\Omega_{\rm DM}\sim 0.22$, 
and cosmological constant $\Omega_\Lambda\sim 0.74$
\cite{spergel07}; therefore, $\sim 96\%$ of the matter and energy
in the Universe is elusive. We describe below how we suppose that these numbers fit in our
picture of the physical world.

\section{Standard solutions }

Extensions of the Standard Model of particle physics have many
DM particle candidates \cite{bertone04}. In supersymmetric models, 
possible DM particles are neutralinos, sneutrinos, gravitinos, axinos. 
Among non-supersymmetric candidates, there are sterile neutrinos,
which are neutrinos that do not have Standard Model weak
interactions; axions, that were introduced to solve the CP problem
in strong interactions; Kaluza-Klein excitations of the Standard Model fields,
like the first excitation of the hypercharge gauge boson, in theories
with extra space dimensions. 

DM particles are supposed to have  
masses in the range between $\sim 1$ GeV and $\sim 100$ TeV, typically, depending
on the various assumptions of the individual model, but scalar particles with masses below
1 GeV can also have the required properties
of DM particles, namely the relic abundance
achieved when they freeze out of thermal equilibrium (this happens when their self-annihilation
rate becomes smaller than the expansion rate of the Universe), the measured $\gamma$-ray 
fluxes expected by the self-annihiliation, and the limits from particle 
physics experiments \cite{boehm04}.
Very massive DM particles, the so-called {\it wimpzillas}, with mass $>10^{10}$ GeV, are also possible
candidates but they must have been out of thermal equilibrium during freeze-out
and thus their relic abundance is independent of their annihilation cross section.
These supermassive particles can be produced gravitationally at the end of inflation.

Search for the DM particle candidates can be separated into direct
and indirect detections \cite{fornengo06}. If the DM particle cross section for
scattering off matter is large enough,
one can hope to measure the nuclear recoils in sensitive 
detectors located underground, to minimize the background noise due to cosmic rays
and radioactive decays from earth rocks. 
Indirect detections rely on the possibilities of detecting
the $\gamma$ photons or the particles (neutrinos, positrons, antiprotons, antideuterons 
\cite{donato07})
produced during the annihilations of DM particles; these annihilations occur in high density
regions of our cosmic surroundings, mainly the Sun, which
is relevant as a source of neutrinos, the Galactic center, and
the DM halo of the Milky Way itself in which the earth is embedded. 
Despite the large effort poured into the many past and present experiments, 
none has yet provided a detection
which can convincingly be interpreted as due to a DM particle (see, e.g.,
\cite{bertone07}). Much hope is
of course put into the data coming from the LHC experiments.

The cosmological constant problem appears to be more serious than 
the DM problem. Einstein equations 
\begin{equation}
R_{\mu\nu}-{1\over 2}g_{\mu\nu}R = {8 \pi G\over c^4}T_{\mu\nu} + {\Lambda\over c^2} g_{\mu\nu} \; ,
\end{equation} 
with the usual meaning of the symbols,
admit the presence of an arbitrary constant $\Lambda$.
The Friedmann solutions with a  positive constant $\Lambda$ fit very satisfactorily 
the observational evidence of an accelerating universe.

The problem arises when one wishes to attach a physical interpretation to 
$\Lambda$. A classical physics approach is to consider the $\Lambda$ term a contribution
to the energy-momentum tensor $T_{\mu\nu}$ and to introduce a dark energy fluid
with equation of state $\rho_\Lambda = - p_\Lambda/c^2 = \Lambda c^2/8\pi G$. Since observations indicate
$\Lambda>0$, the dark energy fluid has negative pressure. Extensions of the equation of state
$p_\Lambda=w \rho_\Lambda$ consider an evolving $w$ (see, e.g.,
\cite{copeland06} for a review), but current
observations suggest $w= -1$ at all probed epoches \cite{astier06}, so models more sophisticated
than a simple constant $\Lambda$ seem unnecessary. It remains to be seen what sets the
value of $\Lambda$. 

In the context of quantum field theory, one interprets the 
energy density $\rho_v(t_0)=\Lambda /8\pi G,$\footnote{I switch to natural units for 
this brief discussion.} at the present time $t_0$, as the ground state of 
the vacuum. In the standard hot Big Bang model, 
the phase transitions occurring in the very early Universe decrease the
vacuum energy density by the quantity $\Delta \rho_v\sim m^4$, where $m$ is 
the mass characteristic of the symmetry break. Specifically,
one finds $\Delta\rho_v^{\rm GUT}\sim 10^{60}$ GeV$^4$, $\Delta\rho_v^{\rm SUSY}\sim 10^{12}$ GeV$^4$,
$\Delta\rho_v^{\rm EW}\sim 10^8$ GeV$^4$, $\Delta\rho_v^{\rm QCD}\sim 10^{-4}$ GeV$^4$ for
the phase transitions which are supposed to occur. Now, we
must have $\rho_v(t_P)=\rho_v(t_0)+\sum\Delta\rho_v$, where $t_P$ is the Planck
time. It follows that $\rho_v(t_P)=(1+10^{-108})\sum\Delta\rho_v$,
because $\rho_v(t_0)=10^{-48}$ GeV$^4$. The $\Lambda$ problem thus
translates into an extreme fine-tuning problem, because $\rho_v(t_P)/\sum\Delta\rho_v$ is
extremely close to $1$ but not exactly $1$. The problems of course would disappear
if $\Lambda $ were exactly zero \cite{weinb89}. 
 
\begin{figure}[t]
        \centerline{\epsfxsize=0.7\hsize \epsffile{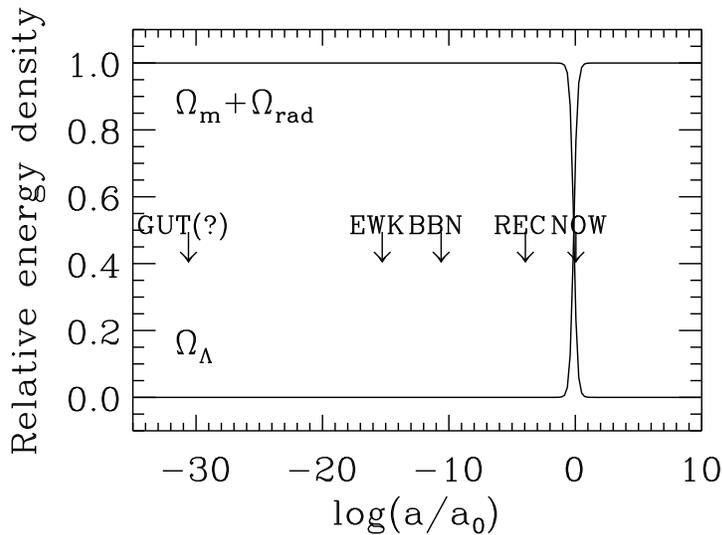}}
        \caption[*]{The relative contribution of the cosmological
constant and matter and radiation to
the energy density budget of the Universe as a function 
of the scale factor $a$. From \cite{kolb07}.
\label{fig:coinc}}
\end{figure}

The $\Lambda $ problem complicates when we consider the time evolution of the matter
and $\Lambda $ contributions to the energy budget of the Universe. Friedmann
equation yields $\Omega_m(t)=\Omega_m(t_0)/E(a)$ and $\Omega_\Lambda(t)=\Omega_\Lambda(t_0)a^3/E(a)$,
where $E(a)=\Omega_m(t_0)+[1-\Omega_m(t_0)-\Omega_\Lambda(t_0)]a + \Omega_\Lambda(t_0)a^3$
and $a(t)$ is the scale factor. 
Figure \ref{fig:coinc} shows
that we live exactly at the transition time between a universe dominated
by $\Omega_m$ and a universe dominated by $\Lambda$. In other words, we
are living at a very special epoch for our Universe and this situation is
somewhat embarassing  unless we resort to the anthropic principle.

\section{Ways out? }

When applied to cosmic structure on galactic and larger scales, GR and its newtonian weak-field 
limit fail at describing the observed phenomenology. To reconcile the 
theory with observations, we need to assume that $\sim 85$\% of the mass  
is seen only through its gravitational effect and
that $\sim 74$\% of the energy content of the Universe
is due either to an arbitrary cosmological constant or to
a not yet well defined dark energy (DE) fluid. The alternative conclusion
we can draw from this failure is that GR is not correct on these 
large cosmic scales. 

An extended/modified theory of gravity is required if we want to unify
quantum field theory with GR. The final theory either
will provide us with a natural explanation for the existence
of DM and DE or will explain the observed phenomenology without
one or either of them. Thus, the scientific community
has considered simplified models hoping that they can be
derived as effective theories from the yet unkown ultimate unification theory.
Over the last 25 years, the number of proposed theories of gravity is enormous
and I will not attempt to list them all here.
I will rather mention a few (random) examples, to provide the reader with a taste of the wealth of 
suggestions appeared in the literature.

\begin{figure}[t]
        \centerline{\epsfxsize=0.5\hsize 
\epsffile{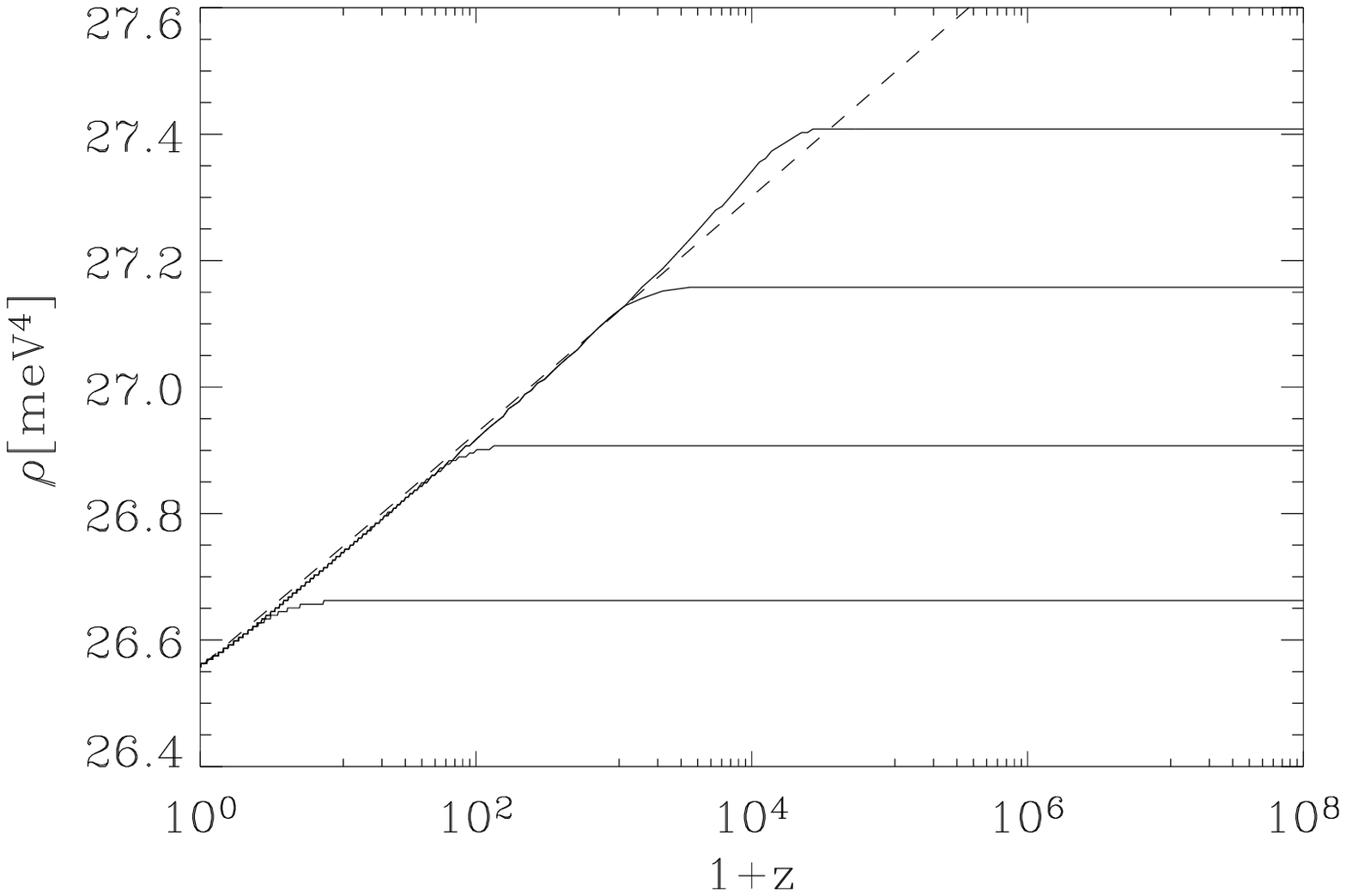} 
\epsfxsize=0.5\hsize\epsffile{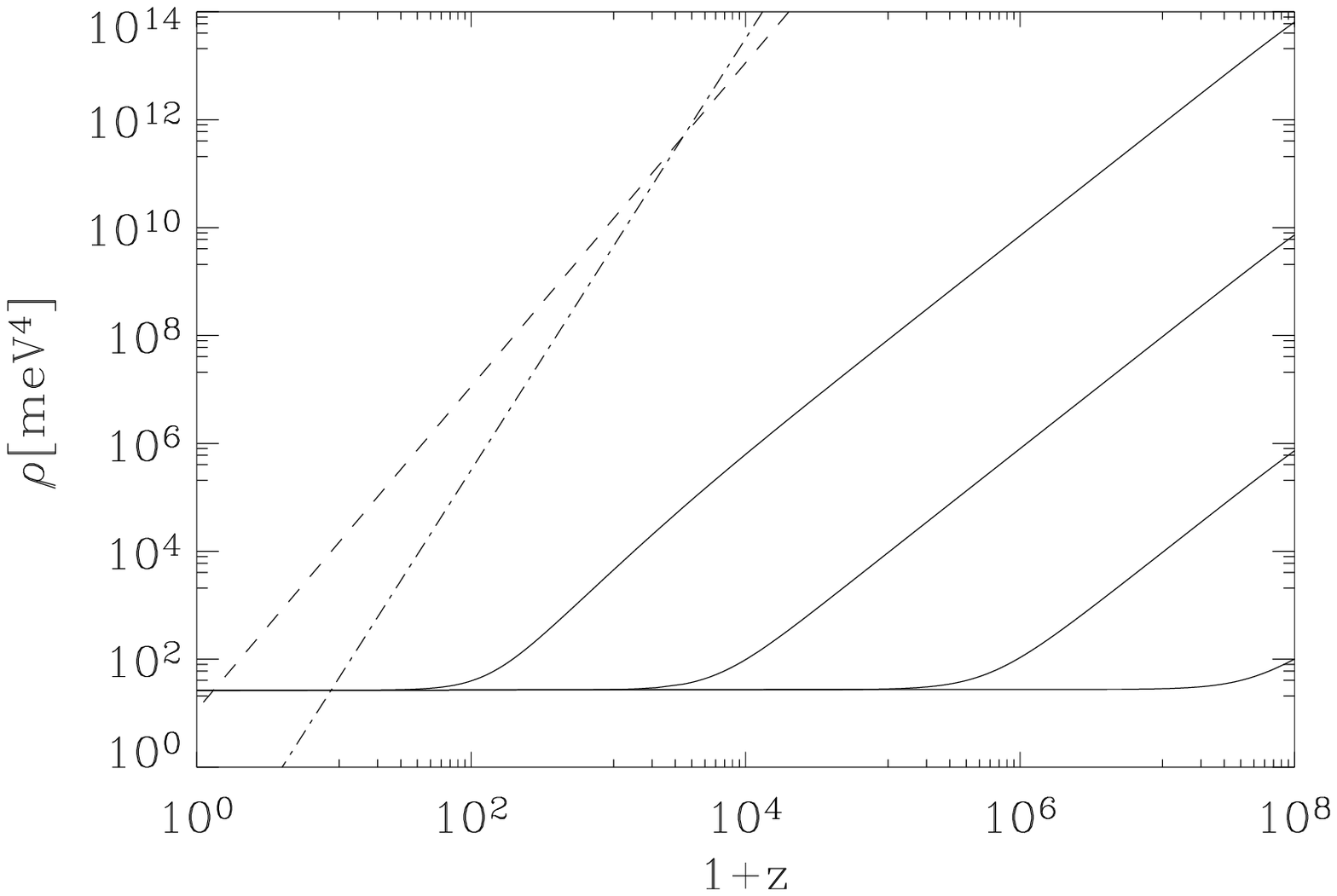}}
        \caption[*]{In the left panel, the solid lines show the time evolution 
of the DE density in quintessence
models with $w=-0.999$ today and different initial values of the scalar field; 
the dashed curve is the
attractor solution. The right panel shows the DE density evolution for non-minimally
coupled models with the same initial values of the scalar field as in
the quintessence models; however, the 
range of initial DE densities they span is much wider than in the quintessence models
because these latter models are forced to have a flat potential $V$ to yield $w=-1$ today. 
The dashed and dot-dashed curves in the right panel show the matter and radiation
energy density evolution. From \cite{matarrese04}. 
\label{fig:scalartens}}
\end{figure}

Consider the gravitational field action
$S \propto \int {\cal L} \sqrt{-g} d^4x$,
with the lagrangian density ${\cal L}=R + \Lambda$ in the case of the Einstein-Hilbert
action. The simplest modification to the lagrangian is to assume
${\cal L}=f(R)$, where $f$ is a generic function (see \cite{capoz07} for
a review). A simple power law
$f(R)\propto R^n$ can describe both the rotation curves of galaxies and
the supernovae Hubble diagram, but not with a power $n$ that admits
a viable matter dominated epoch followed by an acceleration epoch \cite{amendola07}. 

Assuming
the existence of an additional scalar field $\phi$, we can write 
a lagrangian ${\cal L}=f(\phi,R) - \partial^\mu\phi\partial_\mu\phi/2 - V(\phi)$,
where $V(\phi)$ is the scalar field potential.
Models with $f(\phi,R)= R$ are known as quintessence models 
and models with $f(\phi,R)= F(\phi)R$ are known as non-minimally coupled
scalar-tensor theories of gravity (see \cite{fujii03} for a review). Both theories
have a dynamic DE fluid that evolves towards the current value
of $\Omega_\Lambda$. Unlike the quintessence models, the non-minimally 
coupled theories can easily solve 
the fine-tuning problem (Figure \ref{fig:scalartens}) \cite{matarrese04}; 
neither model however solves the coincidence problem. 
 
More drastically, conformal gravity chooses the contraction 
of the Weyl tensor as the lagrangian density \cite{mann06}: ${\cal L}=C_{\mu\nu\lambda\kappa} 
C^{\mu\nu\lambda\kappa}$. This fourth-order theory of gravity \cite{schmidt07} is invariant 
under the conformal transformations $g_{\mu\nu} \to 
\Omega(x)g_{\mu\nu}$, where $\Omega(x)$ is a function of the 
4 space-time coordinates. Conformal gravity is claimed to 
describe both the rotation curves of galaxies and the acceleration
of the universe without DM and DE \cite{mann06}, but it is unable to produce 
enough deuterium during a high-density/high-temperature 
phase of the early universe \cite{elizondo94}, and it 
does not describe correctly the phenomenology of gravitational 
lensing \cite{pireaux04a,pireaux04b} and of clusters of galaxies \cite{horne06}.
 
\begin{figure}[t]
        \centerline{ \epsfxsize=0.6\hsize 
\epsffile{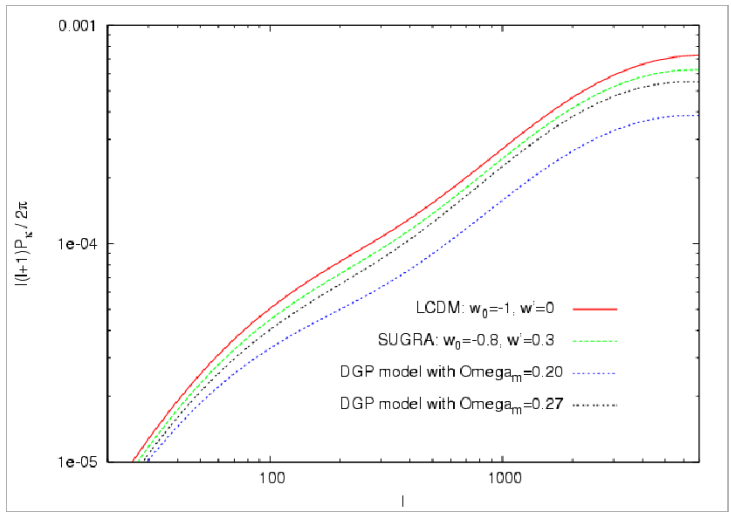}}
        \caption[*]{Weak lensing convergence power spectrum
for the DE models indicated. The two lowest curves are DGP models. 
From \cite{ishak06}.
\label{fig:DGP}}
\end{figure}

Inspired by string theories, braneworld models assume that our 3+1-dimensional 
universe is a mem-{\it brane} embedded in a $3+D+1$-dimensional space-time.
In string theories, the $D$ extra space dimensions are compactified on scales much smaller
than the elementary particle scales. Gravity can leak into these
extra space dimensions, thus explaining the hierarchy problem, i.e. the weakness 
of gravity compared to the other fundamental
forces. In braneworld cosmologies one dimension is
not compactified and can be an infinite {\it bulk}. Gravity also leaks
into this dimension and the attenuation of gravity on very large
scales can be responsible for the late-time acceleration of the universe 
(see \cite{maartens04} for a review).
The gravity attenuation has 
consequences on the formation of cosmic structures. For example, the model
suggested by Dvali, Gabadadze and Porrati (DGP) 
\cite{dvali00,lue06}  yields a weak lensing converging
power spectrum sensibly smaller than the standard $\Lambda$CDM model 
(Figure \ref{fig:DGP}) \cite{ishak06}.

The braneworld models only attempt to solve the DE problem 
and still assume the existence of DM.
The Unified Dark Matter models (see, e.g., \cite{berta07}) 
assume the existence of a single exotic fluid responsible for both
DM, at high density, and DE, at low density. A celebrated example is the Chaplygin
gas \cite{kamen01} that is assumed to have the equation of state $p\propto -\rho^{-\alpha}$.  

All these extended/modified theories of gravity 
have been conceived to describe either the Universe expansion
history or the dynamics of cosmic structure generally in virial equilibrium (mostly
galaxies) or both (see e.g. \cite{capo07}). 
It remains to be seen if the formation and evolution of cosmic
structure can be successfully reproduced in these theories; to this task,
assuming that gravitational instability is the driving process, we need to
calculate the initial field of the density perturbations. Some attempts
towards this direction have been accomplished, for example, for the TeVeS model \cite{skordis08}.

Among the models proposing new physics, the Quasi-Steady State Cosmology (QSSC)
has the longest history and is the most revolutionary (see \cite{narl08} for a review). 
It originates from the classical steady-state model
of Bondi, Gold and Hoyle \cite{bondi48, hoyle48}, the first model to require
(rathen than explain) a negative deceleration parameter $q_0$ \cite{hoyle56}.
The QSSC introduces a scalar field $C$ which is responsible for the
continous creation of matter in active galactic nuclei, rather
in a single big bang event, and for the long-term expansion of the
Universe. The scalar field $C$ enters the action with a negative 
kinetic energy term to compensate for the creation of matter. 
In this model, large-scale structure does not form
by gravitational instability but is determined by the process
of mass ejection from randomly distributed creation centers \cite{nay99}.
The model explains many observables without DM and DE, but it is unclear
if it is able to reproduce in detail the observed evolution 
of galaxy clustering.

\section{Conclusion}

Astronomy has posed relevant problems in physics over the centuries. We are
living at exciting times, where the phenomenology that we see on the sky
fits unsatisfactorily in the laws that, in our mind, rule 
the natural world. To constrain the ideal picture that we wish to formulate
to describe nature, we need to keep extracting information from the sky above
us. Formulating this picture
is a paramount task that does not seem to be close to success yet. But 
the goal is unpredictable, and the path is fascinating. It's worth a try.

\section*{\normalsize \bf Acknowledgements}

\normalsize

I wish to thank Steven Bass for organizing such a stimulating meeting, 
Carlo de los Heros, Chiara Ferrari, Eelco van Kampen, and Sabine Schindler for fruitful
discussion during those days. Special thanks to Chiara for showing me 
the sweet sides of Innsbruck. I thank Steven Bass, Daniele Bertacca, 
Nicolao Fornengo and Margaret Geller for a careful reading 
of the manuscript and suggestions. Support from the institutions sponsoring the meeting, 
the PRIN2006 grant ``Costituenti fondamentali dell'Universo'' of
the Italian Ministry of University and Scientific Research
and from the INFN grant PD51 is gratefully acknowledged.

%\newpage

\end{document}